\documentstyle[12pt,aasms4]{article}

\lefthead{Cimatti et al.}
\righthead{The UV radiation from $z\sim2.5$ radio galaxies}

\begin{document}

\title{The UV radiation from $z\sim2.5$ radio galaxies: Keck spectropolarimetry
of 4C 23.56 and 4C 00.54}

\author{Andrea Cimatti and Sperello di Serego Alighieri}
\affil{Osservatorio Astrofisico di Arcetri, Largo E. Fermi 5, 
I-50125, Firenze, Italy}

\author{Jo\"el Vernet}
\affil{European Southern Observatory, Garching bei M\"unchen, Germany}

\author{Marshall H. Cohen}
\affil{Astronomy Department, California Institute of Technology, USA}

\and

\author{Robert A.E. Fosbury}
\affil{ST-ECF, Garching bei M\"unchen, Germany}

%
%

\begin{abstract}

We present the results of deep spectropolarimetry of two powerful radio
galaxies at $z\sim2.5$ (4C 00.54 and 4C 23.56) obtained with the 
W.M. Keck II 10m telescope, aimed at studying the relative contribution
of the stellar and non-stellar components to the ultraviolet continuum. 
Both galaxies show strong linear polarization of the continuum between 
rest-frame $\sim$1300-2000~\AA, and the orientation of the electric vector is 
perpendicular to the main axis of the UV continuum. In this sense, our 
objects are like most 3C radio galaxies at $z\sim1$. The total flux spectra 
of 4C 00.54 and 4C 23.56 do not show the strong P-Cygni absorption 
features or the photospheric absorption lines expected when the UV 
continuum is dominated by young and massive stars. The only features detected 
can be ascribed to interstellar absorptions by SiII, CII and OI. Our 
results are similar to those for
3C radio galaxies at lower $z$, suggesting that the UV continuum of 
powerful radio galaxies at $z\sim2.5$ is still dominated by non-stellar 
radiation, and that young massive stars do not contribute more than 
$\approx$50\% to the total continuum flux at 1500~\AA.

\end{abstract}


\section{Introduction}

High-$z$ radio galaxies (H$z$RGs) are observable to very high redshifts
and can be used to study the formation and evolution of massive elliptical
galaxies (see McCarthy 1993 for a review). One of the most controversial 
issues is the physical cause of the alignment between the radio 
source and UV continuum axes of the H$z$RGs (the so called `alignment effect', 
Chambers, Miley \& van Breugel 1987, McCarthy et al. 1987). Two main competing
scenarios have been proposed. The first is star formation induced by 
the propagation of the radio source through the ambient gas (see McCarthy 
1993 and references therein); the second  explains the alignment effect as 
the result of a hidden quasar whose radiation is emitted anisotropically 
and scattered towards the observer, producing strong linear polarization 
perpendicular to the radio-UV axis (Tadhunter et al. 1988; di Serego Alighieri 
et al. 1989). The latter scenario is closely related to the unification of 
powerful radio-loud AGN, and provides a way of testing it directly
(see Antonucci 1993 and references therein). After the first detections 
of strong UV polarization in H$z$RGs obtained with 4m-class telescopes 
(di Serego Alighieri et al. 1989; Jannuzi \& Elston 1991;
Tadhunter et al. 1992; Cimatti et al. 
1993), recent observations made with the Keck I 10m telescope have 
demonstrated the presence of spatially extended UV continuum polarization 
and of hidden quasar nuclei in some of the 3C radio galaxies at $0.7<z<1.8$, 
favoring the beaming and scattering scenario (Cohen et al. 1996; Cimatti et al. 
1996,1997; Dey et al. 1996; Tran et al. 1998). On the other hand, Dey et 
al. (1997) have recently shown that the UV continuum of 4C 41.17 ($z=3.8$) 
is unpolarized and consistent with that of a typical starburst galaxy. 
The most stringent comparison between the starburst and the 
scattering scenarios can be performed at $\lambda_{rest} \sim 
1000-2000$~\AA, where most of the strongest spectral features 
of O and B stars are located. This spectral window can be covered 
from the ground by observing radio galaxies at $z>2$. We have started 
a program of observations of these galaxies using spectropolarimetry 
at the Keck II 10m telescope, and in this Letter we report on the first two 
objects we have studied, concentrating on their continuum and absorption
line properties. Throughout this paper we assume 
$H_0=50$ kms$^{-1}$ Mpc$^{-1}$ and $q_0=0$.

\section{Observations and analysis}

The observations were made on UT 1997 July 5-7 with the Keck II 10m 
telescope equipped with the instrument LRIS (Low Resolution Imaging
Spectrometer; Oke et al. 1995) and its polarimeter (Goodrich et al. 1995). 
The LRIS detector is a Tek 2048$^{2}$ CCD with 24$\mu$m pixels which 
correspond to a scale of 0.214 arcsec pix$^{-1}$. We used a 300~line/mm 
grating and a 1.5 arcsecond wide slit, providing a dispersion of 2.4~\AA/pixel.
The spectral resolution, measured from sky and HgKr lamp lines, is $\approx$13~
\AA~ (FWHM). The seeing ranged from about 0.5 to 1.0 arcseconds. 
Polarized (VI Cygni 12) and unpolarized/
spectrophotometric (BD+33$^{\circ}$2642) standard stars (Schmidt, Elston \& 
Lupie 1992) were observed in order to check and calibrate the polarimeter
and to flux calibrate the spectra of the radio galaxies. Details on the observation 
technique and on the data reduction can be found in recent papers (Cimatti et 
al. 1996; Dey et al. 1996; Cohen et al. 1997). The statistical errors on
$P$ and $\theta$ have been treated following the method outlined
by Fosbury, Cimatti \& di Serego Alighieri (1993) and developed by Vernet
et al. (in preparation).

We observed 4C~23.56 ($z=2.482$; Knopp \& Chambers 1997)
and 4C~00.54 ($z=2.366$; R\"ottgering et al. 1997). These galaxies 
were selected to have Ly$\alpha$ redshifted to $\lambda>4000$~\AA~
and to be observable continuously for several hours.
4C~23.56 was observed with the slit oriented at P.A.=47$^{\circ}$ in 
one set $4\times1800$ seconds + one set (1800+1626+1320+1320) seconds + 
one set (2700+$3\times2100$) seconds, corresponding to a total integration 
time of about 6.2 hours. 4C~00.54 was observed with the slit oriented at
P.A.=134$^{\circ}$ in one set $4\times1800$ seconds + one set $4\times2280$ 
seconds (i.e. about 4.5 hours). 4C~23.56 has two main components separated
by about 5 arcseconds (called $a$ and $b$); both have strong Ly$\alpha$ 
emission, but most of the continuum emission comes from the south-west 
region (component $a$) (Knopp \& Chambers 1997). The spectra of 4C 00.54 
and 4C 23.56$b$ were extracted with an aperture of 19 pixels (4.1 
arcsec), whereas an aperture of 23 pixels (4.9 arcsec) was used for
4C 23.56$a$. Since the nights were not photometric, the 
spectra have been scaled to the published $R$-band magnitudes
of the two galaxies (Knopp \& Chambers 1997; R\"ottgering et al. 1997). 
The spectra were finally dereddened for Galactic extinction using
the Burnstein \& Heiles (1982) maps, which provided $E_{B-V}=0.16$
and $E_{B-V}=0.02$ for 4C 23.56 and 4C 00.54 respectively, and
adopting the extinction curve of Cardelli, Clayton \& Mathis (1989)

\section{Results}

The results of spectropolarimetry are displayed in Figures 1,2,3. 
Both galaxies show linearly polarized UV continuum. The degree of 
polarization rises into the blue for 4C~23.56$a$. For 4C~23.56$b$ 
(Fig. 2) we derive an upper limit of $P_{3\sigma} <6.9$\% 
in the range $\lambda_{obs}=4266-7000$~\AA. The signal-to-noise ratio 
for 4C 00.54 is lower, but we detect significant polarization in the 
bluest part of the spectrum, with $P=11.9 \pm 2.6$\% and $P=13.1 \pm 
2.3$\% at $\Delta \lambda_{obs}$=4199-4667~\AA~ and 4745-5160~\AA~ respectively.
If we define two wide bins, we still measure significant polarization
($P=8.9 \pm 1.1$\% and $P=4.2 \pm 1.3$\% at $\Delta \lambda_{obs}$=4199-5495~\AA~
and 5627-7500~\AA~ respectively), suggesting an increase of $P(\lambda)$ 
into the blue also for this galaxy. 
The position angle of the electric vector ($\theta$) 
is approximately perpendicular to the main axis of the UV continuum of both
galaxies (4C 23.56$a$: P.A.$_{UVcont.}\sim90^{\circ}$; Knopp \& 
Chambers 1997; 4C 00.54: P.A.$_{UVcont.}\sim0^{\circ}$; H. R\"ottgering, 
personal communication). Table 1 shows the properties of the emission
lines in the total flux spectra, but the signal-to-noise ratio of the 
present data is insufficient to reach a definitive conclusion about their
polarization. For 4C 23.56$a$ we derive the 
continuum-subtracted degree of polarization of the strongest emission 
lines: $P_{3\sigma}$(Ly$\alpha)<5.0$\% and $P_{3\sigma}$ (CIV)$<8.0$\%. 
For 4C 00.54 we obtain $P_{3\sigma}$(CIV)$< 9.5$\%, whereas we detect a
formally significant polarization for the Ly$\alpha$ line (P=6.0$\pm$0.7\%, 
$\theta=30^{\circ}\pm2.5^{\circ}$). However, we do not regard this Ly$\alpha$ 
polarization as real because the line is heavily affected by cosmic ray 
residuals.

We searched for the spectral signatures of young massive stars, by
looking for P-Cygni profiles (in the lines NV$\lambda$1240,
OV$\lambda$1371, SiIV$\lambda$1400, CIV$\lambda$1549,
HeII$\lambda$1640, NIV$\lambda$1719) and for unambiguously photospheric
absorption lines not affected by nearby emission lines (such as
SiIII$\lambda$1296, SiIII$\lambda$1417, CIII$\lambda$1428,
SV$\lambda$1502). We compared our spectra to that of the starburst
region B1 in the nearby star-forming galaxy NGC 1741 (Conti, Leitherer
\& Vacca 1996), and we also searched using the line lists of Kinney et
al. (1993).  No clear evidence was found for any of the P-Cygni
absorptions (see Figures 4 and 5). For the photospheric
absorptions, after taking into account the spectral resolution,
we obtain the most stringent limits for SiIII$\lambda$1296
and SV$\lambda$1502, where for both lines we obtain
W$_{\lambda}$(rest)$<0.6$~\AA~ in 4C 23.56$a$ and W$_{\lambda}$(rest)$
<0.7$~\AA~ in 4C
00.54. In comparison, NGC 1741B1 has W$_{\lambda} \sim$0.6~\AA~ for 
both lines
(Conti et al. 1996), and 4C 41.17 has
W$_{\lambda}$(rest)(SiIII$\lambda$1296)=0.8$\pm$0.2~\AA~ and W$_{\lambda}$
(rest)
(SV$\lambda$1502)=0.4$\pm$0.1~\AA~ (Dey et al. 1997). Only for
4C 00.54 do we tentatively detect an absorption line: CIII$\lambda$1428
(W$_{\lambda}$(rest)=0.6~\AA, Figure 5), whereas the same line has W$_{\lambda}
$(rest)$<0.6$~\AA~ in 4C 23.56$a$. The CIII$\lambda$1428 line is observed
typically in O stars (Kinney et al. 1993) and it has W$_{\lambda} \sim$0.5~\AA~
in NGC 1741B1 (Conti et al. 1996). However, because of the lack of
other strong O star features in the spectrum of 4C 00.54, we regard the
detection of the CIII$\lambda$1428 line as uncertain.
The only clear and significantly detected absorption 
lines are Si~II$\lambda$1260, Si~II+OI$ \lambda$1303, C~II$\lambda$1335, 
and Si~II$\lambda$1526 (Fig.~4 and 5, Table 2). These lines are generally 
ascribed to interstellar absorption (Kinney et al. 1993; Sahu \& Blades 1997). 
Their rest-frame equivalent widths (typically 1-2~\AA) are, within the errors, 
similar to those detected in 4C 41.17, i.e. larger than the Galactic values 
(Kinney et al. 1993), but somewhat smaller than the typical values 
observed in nearby starburst galaxies ($\sim$2~\AA; Conti et al. 1996). 

\section{Discussion}

Our observations suggest that the UV spectra of 4C 23.56 and 4C 00.54
are not dominated by young massive stars, whereas the strong
perpendicular polarization indicates the presence of a relevant 
scattered continuum, making 4C 23.56 and 4C 00.54 similar to the 
polarized 3C radio galaxies at 0.7$<z<$2. Adopting the prescriptions
of Dickson et al. (1995) and Manzini \& di Serego Alighieri (1996) 
and assuming the average
HeII$\lambda$1640/H$\beta$ ratio (3.18) observed in radio galaxies
(McCarthy 1993), we estimate that the nebular continuum contributes 
only $\sim$8\% and $\sim$13\% to the total flux at 1500~\AA~ for 
4C 23.56$a$ and 4C 00.54 respectively. 

If we assume that all H$z$RGs have an obscured quasar nucleus which 
feeds the powerful radio source and whose light is scattered by dust
and/or electrons, we can interpret the low (or null) polarization
of 4C 41.17 (Dey et al. 1997) as due to dilution of the scattered
radiation by the unpolarized light of young stars. A limit on the 
amount of stellar light in the UV continuum of 4C 23.56$a$ and 4C 
00.54 can be derived by assuming that the observed polarization is 
diluted by the unpolarized stellar and nebular continua. The 
ratio between the stellar light and the total flux at 1500~\AA~ can be 
roughly estimated as $F_{stars}/F_{total}=\{[1-(P_{obs}/P_0)]-\kappa\}$, 
where $P_{obs}$ and $P_0$ are the observed and intrinsic degree of 
polarization and $\kappa$ is the ratio between the nebular and the total 
continuum. Adopting a half-cone opening angle of 45$^{\circ}$ and an angle
of 90$^{\circ}$ between the cone axis and the line of sight, we derive 
$P_0\sim$30\% and $P_0\sim$50\% for dust (Manzini \& di Serego Alighieri 
1996) and electron scattering (Miller, Goodrich \& Mathews 1991) respectively. 
Thus, adopting $P_{obs}$(1500~\AA)=13.1\% and 14.4\% for 4C 00.54 and 4c
23.56$a$ respectively, for dust scattering we obtain that $F_{stars}/F_{total}
\leq$43\% and $\leq$44\% for 4C 00.54 and 4C 23.56$a$ respectively, whereas 
$F_{stars}/ F_{total}$ increases to $\leq$61\% and $\leq$63\% for electron 
scattering. These ratios can be considered upper limits because we do not 
know if the observed scattered light is really diluted by a stellar continuum, 
and they imply that stellar light cannot contribute more than about half 
of the UV continuum at 1500~\AA. 

The properties of 4C 23.56, 4C 00.54 and 4C 41.17 can be interpreted
in a evolutionary scenario where H$z$RGs at $z>3$ have a major episode 
of star formation, and their AGN scattered component is diluted by
the stellar light, but it becomes observable at lower $z$ when the
starburst ceases. However, given the rapid evolution of the UV light 
from a starburst, it is also possible that 4C 41.17 simply represents
a case dominated by the starburst rather than an evolutionary
sequence. Future observation of a complete sample of H$z$RGs will help 
us to understand the nature of the alignment effect and the evolution 
of the host galaxies of powerful radio sources.

\acknowledgments

We are grateful to the referee, Pat McCarthy, for useful suggestions,
to Arjun Dey and Sofia Randich for helpful comments, and to
Ken Chambers for discussions regarding 4C 23.56.
The W.M. Keck Observatory is operated as a scientific
partnership between the California Institute of Technology and the
University of California; it was made possible by the generous
financial support of the W.M. Keck Foundation. RAEF is affiliated
to the Astrophysics Division, Space Science Department, European Space Agency.

%
%
%
\clearpage

\begin{figure}
\plotone{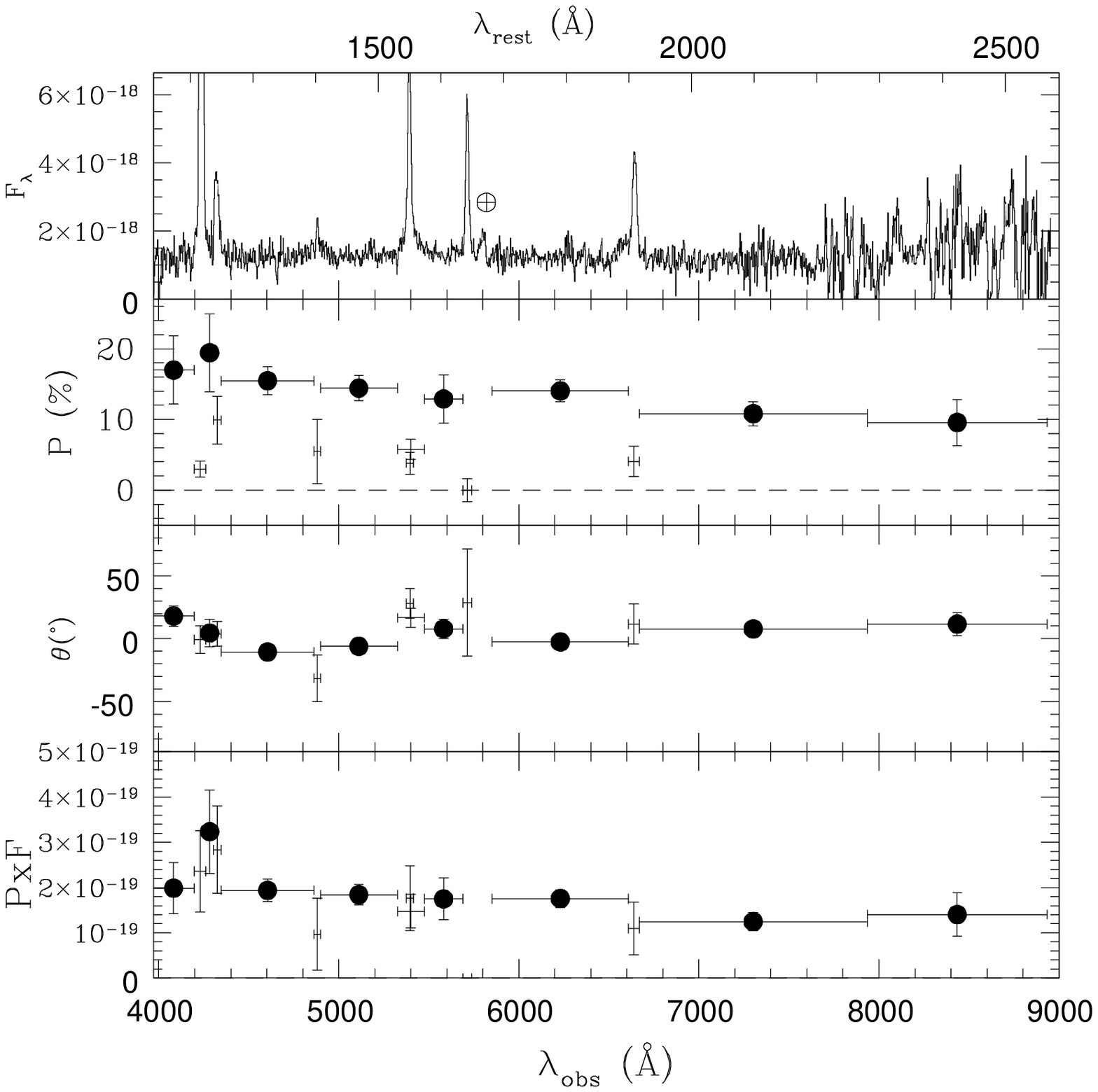}
\caption{
The spectral and polarization properties of 4C~23.56$a$.
From top to bottom: the observed total flux spectrum, the percentage
polarization, the position angle of the electric vector and the
polarized flux spectrum.  Filled circles and crosses indicate
respectively continuum, and emission lines with their underlying
continuum.
}
\end{figure}

\begin{figure}
\plotone{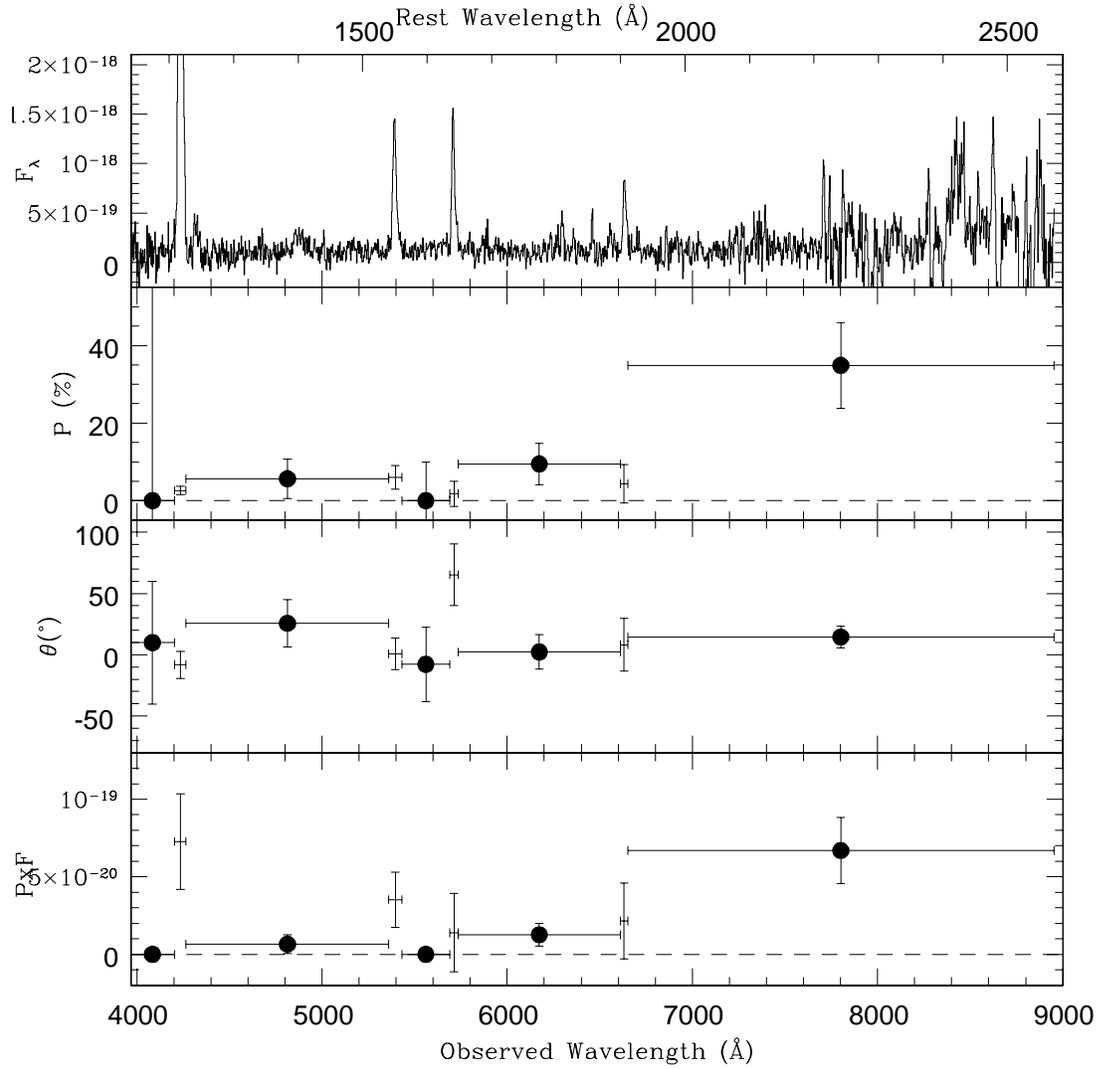}
\caption{
The spectral and polarization properties of 4C~23.56$b$
(same symbols as Figure 1). No significant polarization
is detected.
}
\end{figure}

\begin{figure}
\plotone{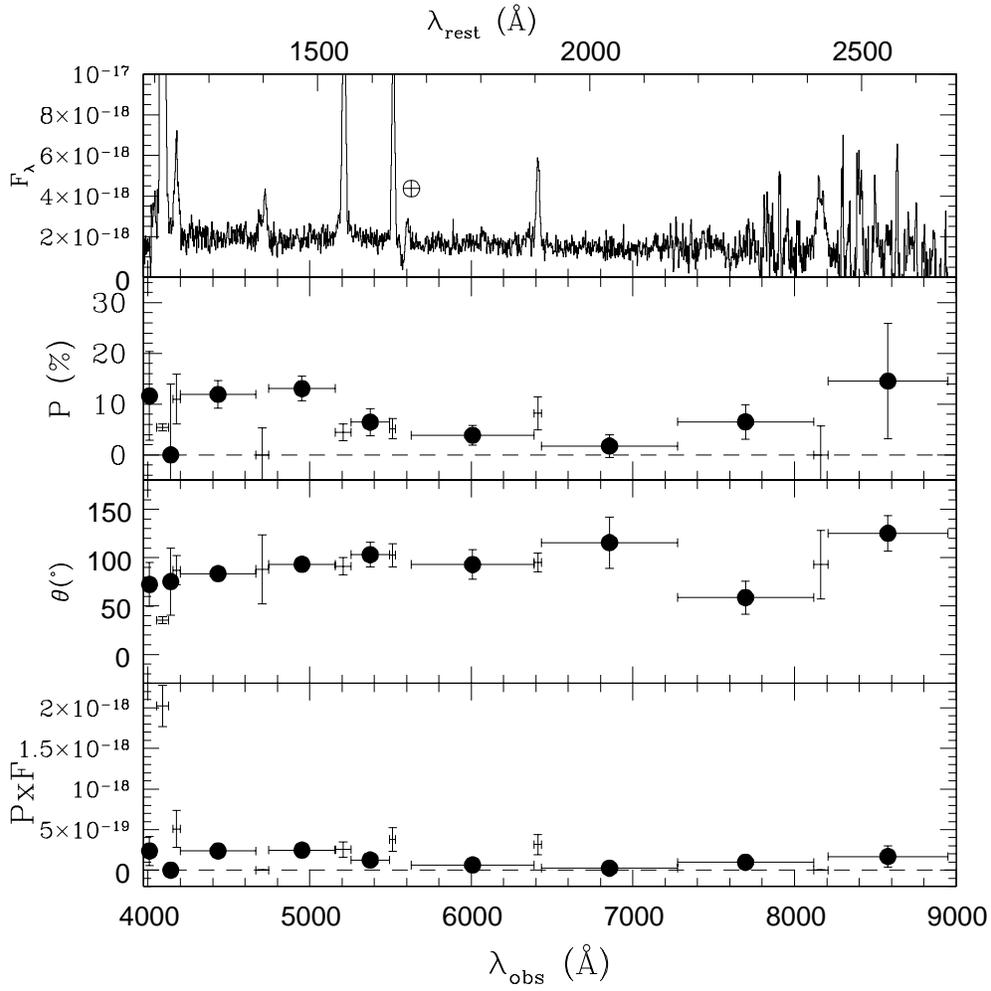}
\caption{
The spectral and polarization properties of 4C~00.54
(same symbols as Figure 1). The polarization of the Ly$\alpha$
line is an artifact due to cosmic ray residuals (see text). 
}
\end{figure}

\begin{figure}
\plotone{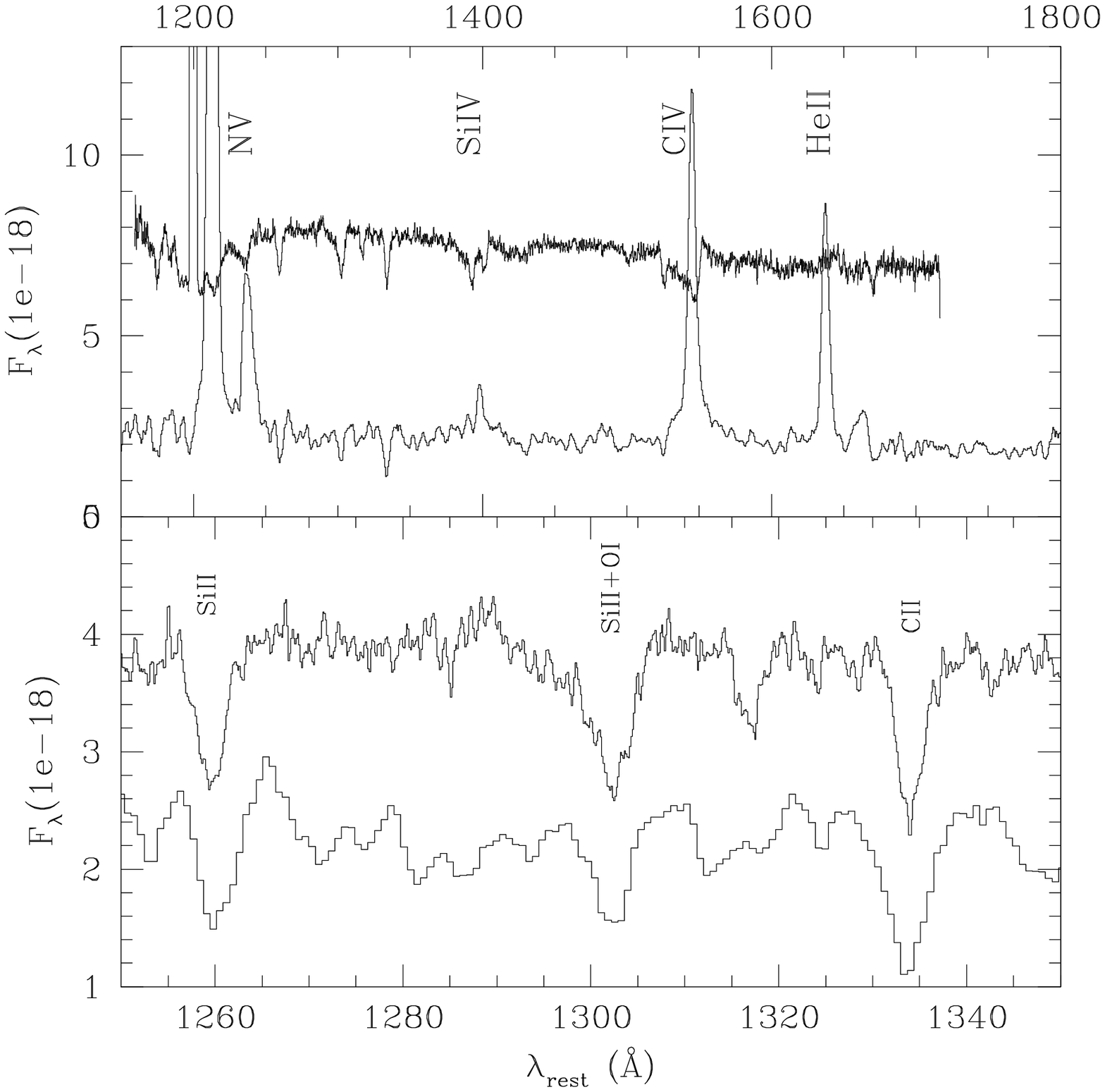}
\caption{
{\it Top panel}. Bottom spectrum : 4C~23.56$a$ corrected for Galactic 
extinction (the spectrum is smoothed with a 3 pixel boxcar). 
Top spectrum : the nearby starburst galaxy NGC 1741B1 scaled to the 
continuum of 4C~23.56$a$ and offset by +5.5. {\it Bottom panel}.
The spectrum of 4C~23.56$a$ (bottom) compared to that of NGC 1741B1
(top, offset by +1.5). 
}
\end{figure}

\begin{figure}
\plotone{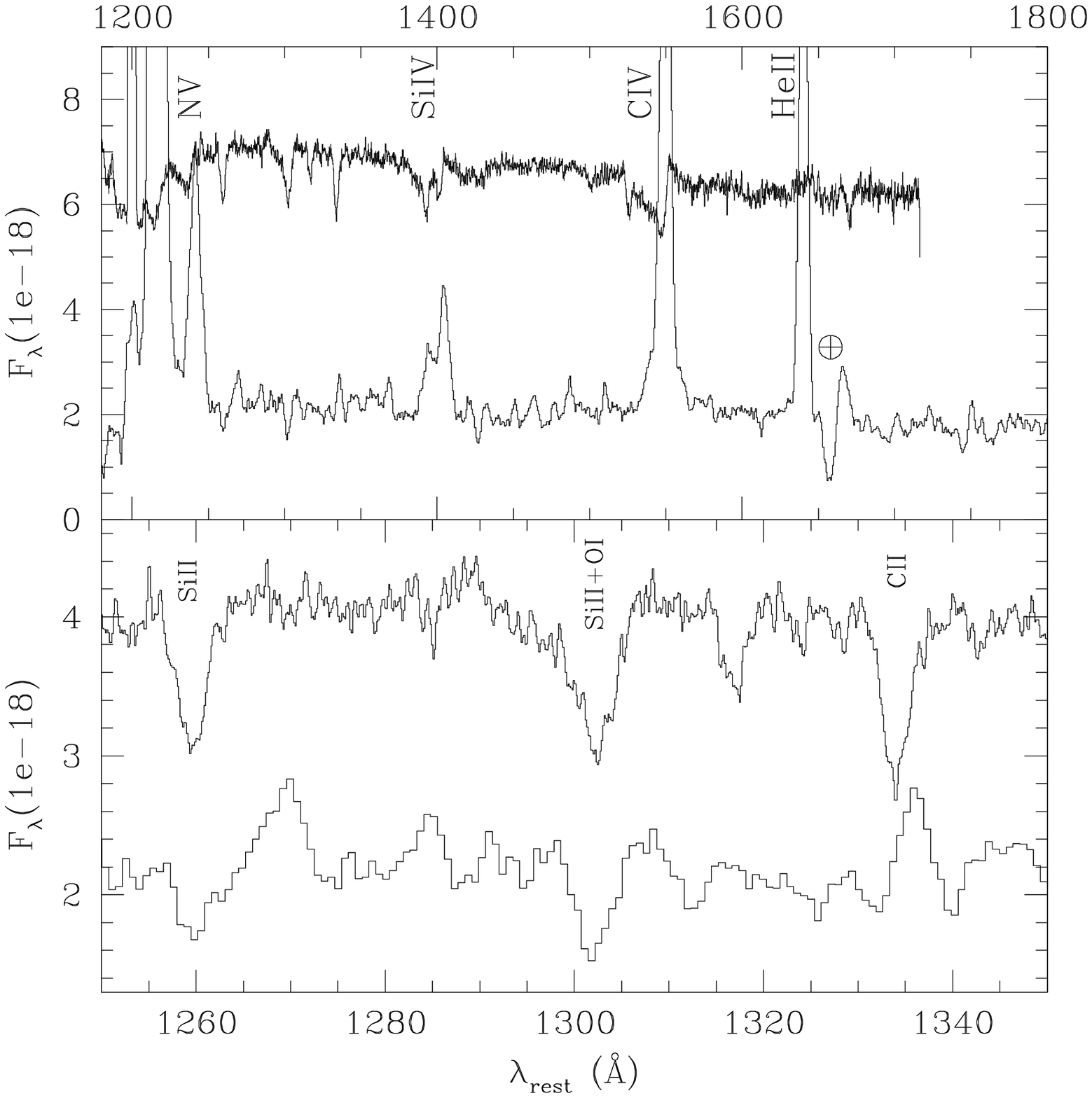}
\caption{
{\it Top panel}. Bottom spectrum : 4C~00.54 corrected for Galactic 
extinction (the spectrum is smoothed with a 3 pixel boxcar). 
Top spectrum : the nearby starburst galaxy NGC 1741B1 scaled to the 
continuum of 4C~00.54 and offset by +5.0. {\it Bottom panel}.
The spectrum of 4C~00.54 (bottom) compared to that of NGC 1741B1
(top, offset by +2.0). 
}
\end{figure}
\clearpage
 
\begin{deluxetable}{llcrr}
\footnotesize
\tablecaption{Emission Line Measurements}
\tablewidth{0pt}
\tablehead{
\colhead{Galaxy} & \colhead{Line} & \colhead{F/F(Ly$\alpha$)} & \colhead{$W_{\lambda}$(obs) (\AA)} & \colhead{FWHM (km s$^{-1}$)} 
} 
\startdata
4C 23.56$a$ & Ly$\alpha$ & 1.00 & 302 & 1458 \nl
 & NV$\lambda$1240 & 0.17 & 58 & 1914 \nl
 & SiIV$\lambda$1394 & 0.02 & 10 & 1400 \\ 
 & SiIV$\lambda$1403 & 0.04 & 15 & 1200 \nl 
 & CIV$\lambda$1549 & 0.26 & 77 & 1057 \nl
 & HeII$\lambda$1640 & 0.19 & 72 & 1024 \nl
 & OIII]$\lambda\lambda$1658-1666 & 0.06 & 29 & 2089 \nl
 & CIII]$\lambda$1908 & 0.16 & 66 & 1219 \nl
 & CII]$\lambda$2327 & 0.10 & 52 & 1607 \nl
 & [NeIV]$\lambda$2424 & 0.17 & 112 & 1954 \nl
4C 23.56$b$ & Ly$\alpha$ & 1.00 & 2200 & 1565 \nl
 & NV$\lambda$1240 & 0.08 & 438 & 2916 \nl
 & CIV$\lambda$1549 & 0.17 & 299 & 1252 \nl
 & HeII$\lambda$1640 & 0.14 & 254 & 972 \nl
 & CIII]$\lambda$1908 & 0.08 & 160 & 922 \nl
4C 00.54 & Ly$\alpha$ & 1.0 & 1488 & 1466 \nl
 & NV$\lambda$1240 & 0.06 & 78 & 2156 \nl
 & SiIV$\lambda$1394 & 0.015 & 22 & 1867 \nl 
 & SiIV$\lambda$1403 & 0.022 & 32 & 1460 \nl 
 & CIV$\lambda$1549 & 0.12 & 139 & 1381 \nl
 & HeII$\lambda$1640 & 0.09 & 129 & 1087 \nl
 & OIII]$\lambda\lambda$1658-1666 & 0.013 & 22 & 1284 \nl
 & CIII]$\lambda$1908 & 0.04 & 74 & 1123 \nl
 & CII]$\lambda$2327 & 0.013 & 38 & 500 \nl
 & [NeIV]$\lambda$2424 & 0.08 & 216 & 2243 \nl
\enddata
 
\tablenotetext{a}{Ly$\alpha$ fluxes: 8.0$\times 10^{-16}$ erg s$^{-1}$
cm$^{-2}$ (4C 23.56$a$), 3.0$\times 10^{-16}$ erg s$^{-1}$ cm$^{-2}$ 
(4C 23.56$b$), 2.8$\times 10^{-15}$ erg s$^{-1}$ cm$^{-2}$ (4C 00.54).
The fluxes are in the observed frame and dereddened using $E_{B-V}$=0.16 
(4C 23.56) and $E_{B-V}$=0.02 (4C 00.54) (see text).
} 
%
%
\end{deluxetable}

\clearpage
 
\begin{deluxetable}{ccccl}
\footnotesize
\tablecaption{Absorption Line Measurements}
\tablewidth{0pt}
\tablehead{
\colhead{Galaxy} & \colhead{$\lambda_{obs}$ (\AA)} & \colhead{$W_{\lambda}$(obs) (\AA)} & 
\colhead{FWHM (km s$^{-1}$)} & \colhead{Identification} 
} 
\startdata
4C 23.56$a$ & 4400.0 & 5.0$\pm$0.7 & 648$\pm$136 & SiII$\lambda$1260.4 \nl
 & 4547.5 & 5.5$\pm$0.6 & 798$\pm$152 & OI$\lambda$1302.2+SiII$\lambda$1304.4 \nl
 & 4657.0 & 8.0$\pm$0.8 & 805$\pm$161 & CII$\lambda$1335 \nl
 & 5323.0 & 3.0$\pm$0.5 & 338$\pm$112 & SiII$\lambda$1527 \nl
4C 00.54 & 4236.5 & 3.0$\pm$0.5 & 389$\pm$141 & SiII$\lambda$1260.4 \nl
 & 4379.3 & 5.5$\pm$0.9 & 959$\pm$198 & OI$\lambda$1302.2+SiII$\lambda$1304.4 \nl
 & 4799.0 & 2.5$\pm$0.5 & 281$\pm$94  & CIII$\lambda$1428 ? \nl 
\enddata
%
%
%
%
\end{deluxetable}

\end{document}